# Early Disease Stage Characterization in Parkinson's Disease from Resting-state fMRI Data Using a Long Short-term Memory Network

Xueqi Guo, *Student Member, IEEE*, Sule Tinaz, and Nicha C. Dvornek

***Abstract*—** Parkinson's disease (PD) is a common and complex neurodegenerative disorder with 5 stages in the Hoehn and Yahr scaling. Given the heterogeneity of PD, it is challenging to classify early stages 1 and 2 and detect brain function alterations. Functional magnetic resonance imaging (fMRI) is a promising tool in revealing functional connectivity (FC) differences and developing biomarkers in PD. Some machine learning approaches like support vector machine and logistic regression have been successfully applied in the early diagnosis of PD using fMRI data, which outperform classifiers based on manually selected morphological features. However, the early-stage characterization in FC changes has not been fully investigated. Given the complexity and non-linearity of fMRI data, we propose the use of a long short-term memory (LSTM) network to characterize the early stages of PD. The study included 84 subjects (56 in stage 2 and 28 in stage 1) from the Parkinson's Progression Markers Initiative (PPMI), the largest-available public PD dataset. Under a repeated 10-fold stratified cross-validation, the LSTM model reached an accuracy of 71.63%, 13.52% higher than the best traditional machine learning method, indicating significantly better robustness and accuracy compared with other machine learning classifiers. We used the learned LSTM model weights to select the top brain regions that contributed to model prediction and performed FC analyses to characterize functional changes with disease stage and motor impairment to gain better insight into the brain mechanisms of PD.

## I. INTRODUCTION

Parkinson's disease (PD) is a common and complex neurodegenerative disorder [1], affecting around 9.4 million people around the world in 2020 [2]. According to the Hoehn and Yahr scaling, five stages of the progression of PD have been proposed [3]. In the early stages, patients only have mild symptoms affecting one side (stage 1) or both sides (stage 2) of the body. Accurate disease staging is crucial for treatment planning, enrollment in clinical trials, and following disease progression.

In recent years, resting-state functional magnetic resonance imaging (rs-fMRI) has been increasingly used to investigate the brain basis of motor and nonmotor symptoms, disease severity, and disease progression in PD [4]. Many of these rs-fMRI studies used the functional connectivity (FC) within and between neural networks as a potential biomarker of PD pathophysiology [5]–[7], but the results have been heterogeneous.

Some studies have investigated machine learning (ML) approaches in PD early diagnosis using rs-fMRI data. Model-based techniques like logistic regression are strongly based on prior statistical assumptions, which may not be applicable for real data with variable dependencies [8]. Model-free algorithms like support vector machines and random forests are able to adapt to inherent characteristics of the dataset with fewer assumptions [8], which outperform traditional model-based classifiers in real-word clinical applications. A support vector machine model trained on randomized logistic regression feature selection was implemented for PD discrimination from connection-wise FC patterns, which reached an accuracy of 80.0% [9]. The support vector machine analysis based on inter group dynamic amplitude of low-frequency fluctuations was reported with significantly increased classification accuracy and abnormal brain activity [10]. The random forest algorithm has been successfully implemented for brain connectivity markers and cognitive impairment [11], [12]. A brain network graph analysis was also proposed on PD diagnosis using rs-fMRI, which achieved an average accuracy of 95% and identified disease associated brain network alterations [13]. However, most current work has mainly focused on distinguishing between PD patients and healthy control subjects [14]–[16], and thus investigating the onset of disease. Further characterization for the early stages of PD, e.g., understanding brain differences in stage 1 and 2, has not been explored, but is necessary to better understand the mechanism and progression of PD.

Recently, deep learning has been successfully implemented in patient representation learning. Convolutional neural networks (CNNs) [17] have been widely applied in medical image analysis cases [18]. A CNN for PD diagnosis from EEG data was reported with high accuracy (88.25%) [19]. A CNN model was trained on structural MRI data to classify PD and healthy controls by transfer learning, achieving an accuracy of 88.9% [20]. However, the huge time and computational consumption required for CNN training on 4D fMRI data would be an obstacle for its maturation in clinical use. Furthermore, the black-box nature of deep learning methods leads to less model interpretability, which is crucial for model utility beyond classification success. Finally, CNNs are well-suited for processing spatial information, but do not take advantage of the temporal sequence of fMRI volumes.

The recurrent neural network (RNN) [21] could capture the temporal dynamics and execute sequential prediction. The long short-term memory (LSTM) [22] unit, a prominent variant of RNN with sophisticated gating mechanisms, is designed to overcome the vanishing gradient problem in long sequences. RNNs and LSTMs have first achieved great success in natural language processing (NLP) tasks [23], while the medical application is also emerging rapidly. In PD diagnosis, the LSTM network has been successfully implemented on voice samples [24] and walking patterns [25]. However, these studies do not investigate the brain functional abnormality. On rs-fMRI datasets, the LSTM model has been investigated on autism identification [26] and Alzheimer's disease prediction [27]. Yet, such work in PD early-stage characterization has not been investigated.

Currently, most rs-fMRI studies in PD enroll a relatively small number ($< 50$) of patients because of the high difficulty of data acquisition. Due to the various data acquisition and pre-processing pipelines among different institutions, as well as the heterogeneity of PD, current research still lacks reproducibility and accuracy across independent datasets. This is certainly a concern for the data-driven approaches. However, some of these concerns can be alleviated with

The manuscript was submitted on YY DDth, 2021. X. Guo was supported by Yale University Biomedical Engineering Ph.D. fellowship. S. Tinaz was supported by the National Institute of Neurological Disorders and Stroke Grant K23NS099478.

X. Guo is with the Department of Biomedical Engineering, Yale University, New Haven, CT, 06511, USA (e-mail: xueqi.guo@yale.edu).

S. Tinaz is with the Department of Neurology, Yale School of Medicine, New Haven, CT, 06519, USA (e-mail: sule.tinaz@yale.edu).

N. C. Dvornek is with the Department of Radiology and Biomedical Imaging and the Department of Biomedical Engineering, Yale University, New Haven, CT, 06511, USA (corresponding author, e-mail: nicha.dvornek@yale.edu).

the landmark open data project in PD called the Parkinson's Progression Markers Initiative (PPMI) [28]. The PPMI has organized the first large-scale and the largest-size public multicenter clinical study to identify PD progression, including advanced imaging, biologic sampling and clinical and behavioral assessments.

In this work, we propose to characterize the early stages of PD from rs-fMRI data on PPMI dataset by applying an LSTM-based model. To the best of our knowledge, this project would be the first use of LSTMs for the identification of PD's early stages using rs-fMRI data. The LSTM model allows for the analysis of the raw time-series data from the brain regions of interest (ROIs), retaining more original imaging information compared to models that predict from the preprocessed FC data. We trained and validated the LSTM model under a 5-times repeated 10-fold stratified cross validation and compared the results with traditional machine learning classifiers that predict from FC measures. We also assess model validity by measuring the association between model output scores and a continuous measure of motor symptom severity in PD. Finally, we interpreted the LSTM model and highlighted the top brain regions and FCs that contributed most to the classification of early stages of PD.

## II. METHODS

### A. Dataset and pre-processing

All subject data were carefully selected and extracted from the public PPMI dataset, noting each subject's sex, age, disease onset side, and disease stage. The original PPMI study was conducted following the Declaration of Helsinki and the Good Clinical Practice (GCP) guidelines approved by the local ethics committees of the 24 participating sites in the US (18), Europe (5), and Australia (1) with informed consent obtained from all the enrolled subjects [28]. A total of 84 age- and sex-matched subjects who were also matched for disease onset side ($p > 0.05$, unpaired two-sample t-test) were selected, including 56 subjects at stage 2 and 28 at stage 1. All rs-fMRI images were acquired for 8.5 min with TR = 2,400 ms, TE = 25 ms, flip angle = 80°, matrix = 68 × 66, and FOV = 222 mm. After the first 4 frames were discarded, each patient's raw rs-fMRI scan contained 206 frames in the 4D sequence (8.24 min), yielding a total of 17,304 frames total in the dataset. Each single frame was saved in nifti format.

Preprocessing was performed using the CONN functional connectivity toolbox v17 [29]. The pre-processing steps included motion correction, outlier detection, normalization to the MNI template, smoothing, ROI extraction, and sequence cropping. Outliers were defined as frame-wise displacement above 0.9 mm or global signal changes above 5 standard deviations. The head motion data of all the subjects are displayed in Table I, and no significant differences in head motion measures between stages were found. Each frame was aligned to the AAL-116 atlas [31] with functional and structural simultaneous gray matter (GM)/white matter (WM)/cerebrospinal fluid (CSF) segmentation and MNI normalization. A Gaussian smoothing at 8 mm full-width at half-maximum (FWHM) was implemented. Denoising steps included correction for physiological and other sources of noise by regressing out the principal components of the white matter and cerebrospinal fluid signal using the CompCor method [30], regression of motion artifacts and outliers, and linear detrending. Finally, data were bandpass-filtered ($0.008 < f < 0.09$ Hz). The mean of the voxel values in each ROI was used as that region's signal. Along the time dimension, each mean ROI time-series was extracted and standardized by dividing by the standard deviation among time frames to represent the relative change.

TABLE I
THE HEAD MOTION DATA OF THE RS-FMRI DATA (MEAN ± STANDARD DEVIATION).

| Stage | Maximum Motion (mm) | Mean Motion (mm) | Outlier Scans |
|---|---|---|---|
| HY1 | 1.2261 ± 0.8207 | 0.2481 ± 0.1004 | 5.3928 ± 5.9587 |
| HY2 | 1.0343 ± 0.8539 | 0.2342 ± 0.1039 | 3.8392 ± 9.0549 |

Considering the relatively small number of early stage subjects for deep learning training, data augmentation was introduced to help the model generalize better and prevent overfitting. The input time sequences were cropped with a fixed sequence length w = 50 (representing 2 min of imaging) and stride length s = 1 to move along the time dimension of the rs-fMRI series. Thus for each subject, 156 cropped sequences were acquired, which boosted the sample size to 13,104 in total.

### B. LSTM Model

We aimed to investigate an LSTM model for PD early stage characterization from rs-fMRI data. We hypothesized that the early stage classification performance would improve compared with traditional ML classifiers that predict from FC measures. The overall workflow of our project is shown in Fig. 1.

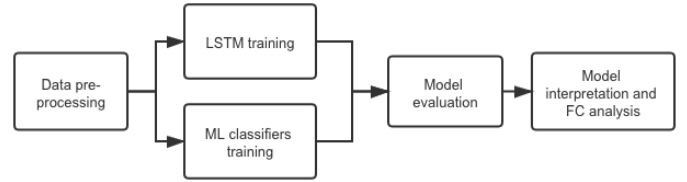

Fig. 1. The overall workflow of our project.

LSTMs are a special type of RNN that are able to address the vanishing gradient and limited long-term memory problems in a vanilla RNN model, taking the previous information and the current data input to update the cell state and hidden state. The key equations in an LSTM cell are:

$$i_t = \sigma\left(W_i x_t + U_i h_{t-1} + b_i\right) \quad (1)$$

$$f_t = \sigma\left(W_f x_t + U_f h_{t-1} + b_f\right) \quad (2)$$

$$\tilde{c}_t = \tanh\left(W_c x_t + U_c h_{t-1} + b_c\right) \quad (3)$$

$$c_t = i_t * \tilde{c}_t + f_t * c_{t-1} \quad (4)$$

$$o_t = \sigma\left(W_o x_t + U_o h_{t-1} + b_o\right) \quad (5)$$

$$h_t = o_t * \tanh\left(c_t\right) \quad (6)$$

where at time step $t$, $x_t \in \mathbb{R}^N$ is the vector of $N$ ROI values, $h_t \in \mathbb{R}^M$ is the hidden state, $c_t \in \mathbb{R}^M$ is the cell state, with an input gate $i_t \in \mathbb{R}^M$ deciding what information from the current estimated cell state is updated, a forget gate $f_t \in \mathbb{R}^M$ deciding how much of the previous hidden state should be discarded, and an output gate $o_t \in \mathbb{R}^M$ filtering the cell state to update the hidden state. $W \in \mathbb{R}^{M \times N}$ is the weights applied to the input, $U \in \mathbb{R}^{M \times M}$

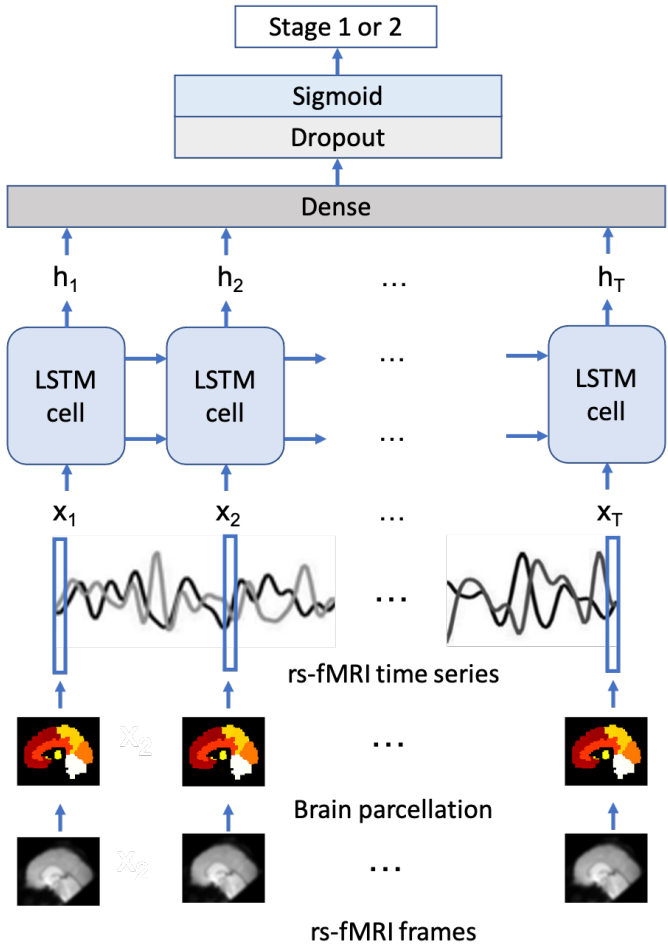

Fig. 2. The proposed LSTM model structure.

### C. Model training and evaluation

To comprehensively evaluate the model performances, 5 repeats of 10-fold stratified cross validation was implemented. For each repeated run, 10% of subjects are selected as test set, 10% are selected as validation set, and the remaining 80% are the training set. Note that the random split was carried out on both stage 1 and stage 2 in order to stratify the imbalanced dataset and keep the ratio of the two stages roughly the same in all the splits. To handle the class imbalance, sample weights are assigned inversely proportional to the stage ratio in the dataset. For the LSTM model and all the ML models, the hyperparameter tuning was first performed by using the validation set to evaluate the performance of the models trained on the training set under the different hyperparameters. After the best hyperparameters were selected, the final model was trained on both the training and the validation set and then evaluated on the test set to best utilize the relatively smaller dataset.

The initial hidden and cell states of the LSTM were set as all zeros. The hidden size $M$ of the LSTM layer was tuned from the list of [16, 32, 64, 128]. To prevent overfitting on the LSTM model, L2 regularization and the early-stopping mechanism were introduced to enhance the model's generalization ability. Training was terminated when the validation loss had not reduced in 10 epochs or when the maximum epoch number 50 was reached. The LSTM network was trained using the cross entropy loss function, Adam optimizer with learning rate = 1e-4, batch size = 200, dropout rate fixed at 0.5, the $\lambda$ of the L2 regularization at 0.01, and other default parameter settings. The LSTM model was implemented using PyTorch and trained on a NVIDIA GeForce GTX 1080 GPU at the Farnam Cluster of the Yale Center for Research Computing.

All the other ML models were implemented via Scikit-learn. The regularization parameter inversely proportional to the regularization strength of LR was searched in the range of [1e-9, 10] and the search range of SVC was [0.01, 10], both with step size as 1e-1 in log scale. For RF, the number of trees were tuned from the list of [10, 50, 100, 200, 500]. The maximum depth of the tree was searched from 2 to 6. The minimum number of samples to split an internal node was selected from the list of [10, 20, 50, 100, 200, 500, 1000, 2000, 3000] and the searching list of the minimum number of samples in a leaf node was [10, 50, 100, 200, 300, 500]. All the other hyperparameters were set as default.

All the evaluation metrics were based on subject-level outputs, which are given by the majority vote of all the sequences from one subject. Evaluation at the subject level better matches the real clinical diagnosis of one label per subject, and the ensembling also has improved performance compared to the sample-wise results. The subject-wise accuracy, precision, recall (sensitivity), specificity, and F1 score were reported. The equations for accuracy, precision, recall, specificity, and F1 score calculation are as below,

$$Accuracy = \frac{TP + TN}{TP + TN + FP + FN} \quad (7)$$

$$Precision = \frac{TP}{TP + FP} \quad (8)$$

$$Recall = \frac{TP}{TP + FN} \quad (9)$$

$$Specificity = \frac{TN}{FP + TN} \quad (10)$$

$$F1 = \frac{2 * Precision * Recall}{Precision + Recall} \quad (11)$$

is the weights applied to the previous hidden state, $b \in \mathbb{R}^M$ is the bias, and $\sigma$ is the sigmoid activation function.

The proposed LSTM model structure is shown in Fig. 2. The model takes the average time-series of ROIs as the input and then utilizes the output of each time step, which aggregates the decoded hidden state of each cell as the input of the fully connected (dense) layer with 1 node. A dropout layer was integrated between the dense layer and the sigmoid activation layer to prevent overfitting, and the final output of the sigmoid layer would be interpreted as the probability of being assigned as each stage, where higher probabilities correspond to higher likelihood of stage 2. This architecture would directly take the signal at every time point into consideration, which would improve the network's robustness handling noisy rs-fMRI data.

We also conducted traditional ROI-based machine learning methods in rs-fMRI analysis as the baseline. A standard pipeline is calculating the FC matrix, representing the correlation or covariance between each ROI pair (connectome), and then using the FC matrices as predictors in a traditional machine learning classifier [33]. Here, we tested random forest (RF), linear support vector machine classifier (SVC) and logistic regression (LR). The FC matrices were calculated using the Ledoit Wolf estimator [34] for large covariance matrices, and only the upper triangle of the matrices were used as the inputs of the ML classifiers to prevent redundancy since the FC matrices are symmetric.

where positive and negative refer to stage 2 and stage 1, respectively, $TP$ is the true positive, $TN$ is the true negative, $FP$ is the false positive, and $FN$ is the false negative.

The corrected repeated k-fold cross validation test [35] was conducted as the significance test for model performance comparison. For a $r$-times $k$-fold cross-validation, the following statistic is calculated,

$$t = \frac{\sum_{i=1}^{k}\sum_{j=1}^{r} x_{ij}}{(k*r)\sqrt{\left(\frac{1}{k*r} + \frac{n_2}{n_1}\right)\hat{\sigma}^2}} \quad (12)$$

where $x_{ij}$ is the difference in the statistic of interest between two models being compared from the $i$th fold of the $j$th cross-validation run, $n_1$ is the number of subjects used for training, $n_2$ is the number of subjects used for testing, and $\hat{\sigma}^2$ is the estimated variance of the differences $x_{ij}$. This test corrects the estimate of the variance by taking the dependency between cross-validation samples into account. The significance level $\alpha$ was set at 0.05.

The Movement Disorders Society-Unified Parkinson's Disease Rating Scale (MDS-UPDRS) is the standard assessment tool for disease severity and progression of PD [36]. The MDS-UPDRS part III motor exam score rates the severity of motor impairment. Higher scores indicate worse motor impairment. We computed the Pearson correlation between the model output scores and the MDS-UPDRS-III scores to evaluate whether the stage classification model output is associated with a more continuous measure of disease severity.

### D. Model interpretation

We interpreted the LSTM model by exploring the learnable input-hidden weights $W$ of the LSTM cell. After z-score normalization, we highlighted the ROIs with the magnitudes of the associated weights above mean and one standard deviation of the weights. These ROIs were considered as the important ROIs for early stage PD classification.

Pairwise FC analysis was then conducted based on the selected top ROIs. The covariance matrices for all subjects were computed from the top ROIs. Then for each edge, the Welch's t-test [37] was used to compare the FC for the ROI pair between the stage 1 and stage 2 subjects, taking into consideration the different number of subjects in each stage. The significance of the Welch's t-statistic was assessed using a permutation test with 10,000 random permutations of the subject stage labels to find whether there was a significant difference in the FC between the two stages. Correction for multiple comparisons was performed by controlling the false discovery rate (FDR) [38].

Regression analysis was performed to analyze the association between the FC of top ROIs and the MDS-UPDRS-III scores. The elastic net model [39] was used to predict MDS-UPDRS-III scores from the pairwise FC between top ROIs. The elastic net regularization parameters were searched using the repeated cross-validation splitting strategy (3 runs, 10 folds), and the optimal hyperparameters were then applied to the model for the entire dataset. The searching range of both $\alpha$ and l1 ratio is [0, 1]. The selected regularization parameters for LSTM were $\alpha$ = 0.1936 and l1 ratio = 0.1000.

The permutation test with 10,000 runs was again used to assess significance of the regression coefficients, where the subject MDS-UPDRS-III scores were randomly shuffled. The top ROI analysis results were also compared with results of similar whole-brain FC analysis to see whether the top ROIs play a dominant part in disease stage progression. All the FC differences were visualized via BrainNet Viewer [40].

## III. RESULTS

### A. Characterization results

Table II summarizes the characterization results of the LSTM model and all the ML classifiers. All the models have been selected with the best hyperparameters. The proposed LSTM model yielded the highest values among all the quantitative metrics, which outperformed all the other ML methods. Under the corrected repeated k-fold cross validation test ($k$ = 10, $r$ = 5, $n_2$ = 8, $n_1$ = 76, degree of freedom = 49, and $t_{thre}$ = -2.010), the results of the LSTM model showed significant improvement in accuracy, F1, and recall compared to all other models. In precision and specificity, the LSTM model showed significant improvement compared to SVC. This suggests the LSTM model has the potential to better extract and utilize temporal information from the rs-fMRI data.

Table III presents the correlation between the model output scores and the MDS-UPDRS-III scores. The SVC model, which resulted in the second highest precision and specificity for stage classification (Table II), did not produce significant correlation of model output and MDS-UPDRS-III scores ($r$ = 0.1748, $p$ = 0.12). The LSTM model showed the highest correlation between the stage prediction score with MDS-UPDRS-III scores ($r$ = 0.4270, $p$ = 0.00003). A high correlation is desired, as the degree of motor impairment plays an important part in early stage distinction. Thus, the proposed LSTM model not only resulted in the best stage classification performance, but also the confidence of the LSTM model's classification produced the highest correlation with a closely related continuous rating of disease severity.

### B. Brain abnormality detection

We detected the top ROIs related to the brain abnormality in disease development by investigating the learnable weights in the LSTM model. Table IV displays the top ROIs with the greatest absolute weights for the overall LSTM model in descending order. The level is the computed z-score of the absolute value of the extracted weights, showing how much the region magnitude is above the mean weight magnitude of all the ROIs.

A diverse set of ROIs represented disease severity. Most notably, the basal ganglia structures including the lenticular nucleus and caudate, and the supplementary motor area together with the postcentral gyrus are implicated in sensorimotor impairment in PD. The cerebellum is involved in tremor generation. The amygdala and insula are limbic structures involved in emotional processing and play a role in anxiety and depression in PD. The calcarine sulcus and superior occipital gyrus are visual processing areas implicated in visual symptoms of PD such as hallucinations. The posterior cingulate is the major hub in the default mode network that shows abnormal FC in PD and other neurodegenerative disorders. Finally, the frontal regions mediate higher cognitive and executive functions and are implicated in cognitive dysfunction in PD even in the early stages and in the absence of dementia [4]. Thus, the top influential ROIs for the LSTM model that play an important role in distinguishing early-stage PD are also relevant brain regions linked to motor and nonmotor functions that are affected in PD.

### C. Brain connectivity analysis

The brain connectivity analysis was carried out first by assessing FC differences between stages 1 and 2 for all the pairwise connections between the top ROIs by using the permutation Welch's t-test. The edges that showed significant differences between the two stages are listed in Table V, and the nodes and edges are displayed in Figure 3. All listed edges were weaker in stage 2 than in stage 1, indicating that a more advanced disease stage in PD is associated with decreased functional connectivity. Interestingly, the influential edges are mostly between the nodes related to nonmotor brain functions. This finding has important clinical implications suggesting that progression even in the early stages of the disease involves FC changes in nonmotor networks, underlining the importance of evaluating the severity of not only motor but also nonmotor impairment in clinical progression studies and prediction models.

TABLE II
EARLY-STAGE PD CHARACTERIZATION RESULTS (MEAN ± STANDARD DEVIATION).

|      | Accuracy | F1 | Precision | Recall (Sensitivity) | Specificity |
|------|----------|-----|-----------|----------------------|-------------|
| LR   | 0.5664 ± 0.1897 | 0.6346 ± 0.2032 | 0.6993 ± 0.2089 | 0.6173 ± 0.2402 | 0.4666 ± 0.3605 |
| RF   | 0.5862 ± 0.1510 | 0.6745 ± 0.1341 | 0.7199 ± 0.1548 | 0.6580 ± 0.1702 | 0.4433 ± 0.3155 |
| SVC  | 0.5811 ± 0.1451 | 0.6508 ± 0.1721 | 0.7216 ± 0.1930 | 0.6373 ± 0.2253 | 0.4766 ± 0.3496 |
| **LSTM** | **0.7163 ± 0.1318** | **0.7912 ± 0.1050** | **0.7794 ± 0.1047** | **0.8226 ± 0.1579** | **0.5833 ± 0.2948** |

TABLE III
THE CORRELATION BETWEEN THE MODEL OUTPUT SCORES AND THE MDS-UPDRS-III SCORES (MEAN ± STANDARD DEVIATION OF 5 RUNS).

|      | Pearson correlation |
|------|---------------------|
| LR   | 0.3212 ± 0.0550 |
| RF   | 0.3911 ± 0.0231 |
| SVC  | 0.1748 ± 0.0455 |
| **LSTM** | **0.4270 ± 0.0595** |

TABLE IV
THE TOP ROIS AND THE ASSOCIATED BRAIN FUNCTIONS WITH THE GREATEST WEIGHT MAGNITUDES FOR THE OVERALL LSTM MODEL.

| Region | Level | Function |
|--------|-------|----------|
| Vermis_10 | 2.4741 | Motor functions |
| Inferior frontal gyrus, orbital part, right | 2.2505 | Higher cognitive functions |
| Calcarine sulcus, right | 1.9351 | Visual functions |
| Middle frontal gyrus, orbital part, left | 1.8824 | Higher cognitive functions |
| Insula, right | 1.8568 | Emotional functions |
| Calcarine sulcus, left | 1.8407 | Visual functions |
| Middle frontal gyrus, orbital part, right | 1.8824 | Higher cognitive functions |
| Caudate nucleus, right | 1.7248 | Motor functions |
| Superior occipital gyrus, right | 1.5491 | Visual functions |
| Superior frontal gyrus, medial, left | 1.5404 | Higher cognitive functions |
| Amygdala, left | 1.5041 | Emotional functions |
| Postcentral gyrus, right | 1.4804 | Somatosensory functions |
| Cerebellum_3, left | 1.4733 | Motor functions |
| Posterior cingulate gyrus, right | 1.4275 | Default mode functions |
| Supplementary motor area, left | 1.3535 | Motor functions |
| Cerebellum_7b, right | 1.4733 | Motor functions |
| Lenticular nucleus (putamen and globus pallidus), right | 1.1469 | Motor functions |
| Cerebellum_6, right | 1.4733 | Motor functions |
| Superior occipital gyrus, left | 1.0606 | Visual functions |

TABLE V
THE SIGNIFICANTLY DIFFERENT EDGES IN STAGE 1 AND 2 BY PERMUTATION WELCH'S T-TEST FOR THE LSTM MODEL.

| Edge | P-value |
|------|---------|
| Postcentral gyrus, right-Superior occipital gyrus, left | 0.0022 |
| Calcarine sulcus, right-Superior occipital gyrus, left | 0.0077 |
| Inferior frontal gyrus, orbital part, right-Middle frontal gyrus, orbital part, right | 0.0093 |
| Superior occipital gyrus, right-Postcentral gyrus, right | 0.0096 |
| Middle frontal gyrus, orbital part, right-Postcentral gyrus, right | 0.0159 |
| Calcarine sulcus, right-Postcentral gyrus, right | 0.0201 |
| Middle frontal gyrus, orbital part, left-Postcentral gyrus, right | 0.0275 |
| Calcarine sulcus, right-Middle frontal gyrus, orbital part, left | 0.0287 |
| Calcarine sulcus, left-Postcentral gyrus, right | 0.0310 |
| Middle frontal gyrus, orbital part, left-Calcarine sulcus, left | 0.0327 |
| Middle frontal gyrus, orbital part, right-Superior occipital gyrus, right | 0.0329 |

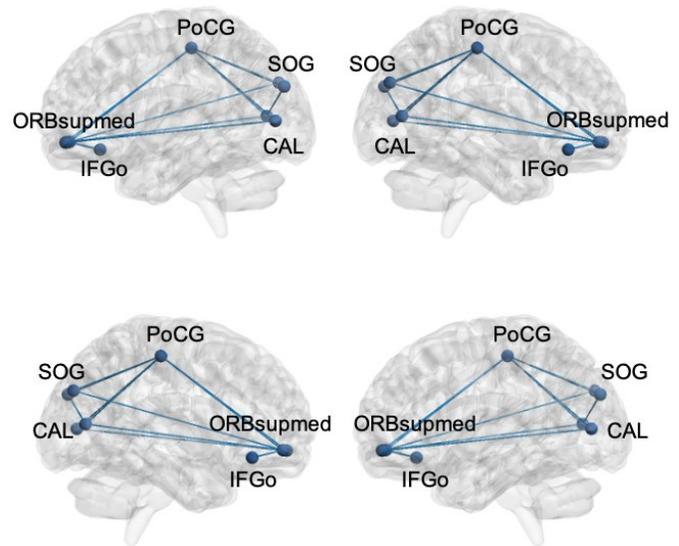

Fig. 3. The significantly different edges in stage 1 and 2 by permutation Welch's t-test for the LSTM model. PoCG: Postcentral gyrus; SOG: Superior occipital gyrus; CAL: Calcarine sulcus; ORBsupmed: Orbitofrontal cortex, superior medial part; IFGo: Inferior frontal gyrus, orbital part.

For motor-related connectivity analysis, the elastic net regression was used to regress the MDS-UPDRS-III score on the FC edges.

Permutation testing was conducted to assess the significance of regression coefficients, where the p-values are calculated as the percentage of permutation results with a coefficient magnitude greater than the magnitude of the original observation. Table VI summarizes the significant edges of the regression results and the related direction of association with motor score. The nodes and edges are displayed in Figure 4. After applying the FDR correction with a false discovery rate of 0.2, the top 3 edges remained significant. Higher MDS-UPDRS-III scores indicate worse motor impairment. Therefore, the "increased" direction denotes a positive relationship between edge strength and motor impairment, whereas "decreased" shows the opposite relationship. The edge strength between the cerebellar vermis and sensorimotor (supplementary motor area and postcentral gyrus) and visual areas (superior occipital gyrus) is associated with worse motor impairment suggesting a compensatory reorganization of brain circuits. Similar compensatory shifts from the defective basal ganglia circuits to the cerebellar circuits have been reported in PD [4].

TABLE VI
THE EDGE WITH SIGNIFICANT WEIGHTS BY PERMUTATION TEST OF THE ELASTIC NET REGRESSION OF MDS-UPDRS-III SCORES FOR THE LSTM MODEL.

| Edge | P-value | Direction |
| --- | --- | --- |
| Calcarine sulcus, left- Postcentral gyrus, right | 0.0004 | Decreased |
| Calcarine sulcus, right- Postcentral gyrus, right | 0.0005 | Decreased |
| Superior occipital gyrus, right- Postcentral gyrus, right | 0.0019 | Decreased |
| Superior occipital gyrus, left- Postcentral gyrus, right | 0.0089 | Decreased |
| Calcarine sulcus, left- Calcarine sulcus, right | 0.0106 | Decreased |
| Vermis_10- Superior occipital gyrus, left | 0.0217 | Increased |
| Middle frontal gyrus, orbital part, right- Superior occipital gyrus, right | 0.0233 | Decreased |
| Posterior cingulate gyrus, right- Cerebellum_6, right | 0.0234 | Decreased |
| Vermis_10-Postcentral gyrus, right | 0.0238 | Increased |
| Inferior frontal gyrus, orbital part, right- Postcentral gyrus, right | 0.0243 | Decreased |
| Middle frontal gyrus, orbital part, right- Cerebellum_6, right | 0.0258 | Decreased |
| Vermis_10- Supplementary motor area, left | 0.0325 | Increased |
| Calcarine sulcus, right- Superior occipital gyrus, left | 0.0434 | Decreased |

A conventional whole-brain FC permutation Welch's t-test and MDS-UPDRS-III score regression analysis were conducted as a comparison with the sub ROI group analysis (details in Appendix). The whole-brain results are visualized in Figure A1 and Figure A2 respectively. Note that none of the edges detected by whole-brain analysis survived the FDR correction, potentially indicating that the traditional whole-brain analysis results in overfitting and thus may not be reliable in revealing FC changes and disease mechanisms that generalize to the greater PD population.

## IV. Discussion

In this work, we investigated an LSTM model for early-stage PD characterization using rs-fMRI data. Under a 5-run, 10-fold repeated stratified cross-validation, the proposed LSTM model performed significantly better than the other traditional ML methods for the classification of stage 1 and stage 2 subjects. The model output scores were also better correlated with the motor severity scale. The learnable weights in the well-trained LSTM model gained meaningful interpretations. The post hoc FC analysis revealed edges that differed significantly between stages 1 and 2 in both classification and regression analyses. The findings of potentially influential top brain regions and abnormal FC among them could provide a deeper understanding of the neuroanatomical substrates of early disease stages in PD.

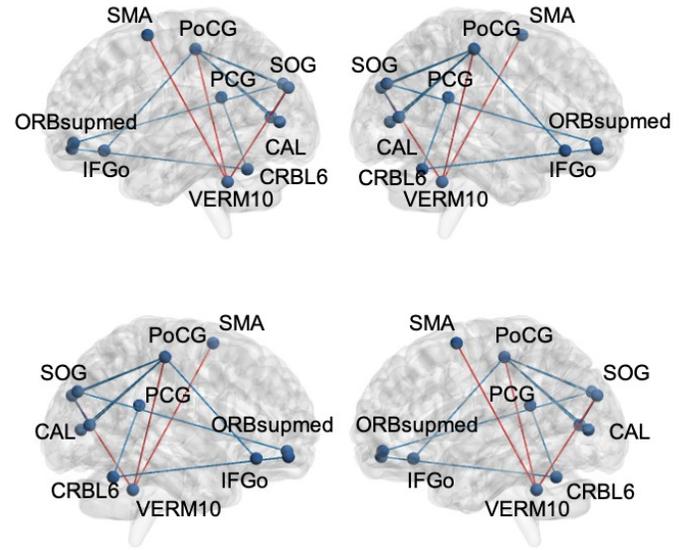

Fig. 4. The edges with significant weights by permutation t-test of elastic net regression of MDS-UPDRS-III score for the LSTM model. Red: edges with positive coefficients. Blue: edges with negative coefficients. SMA: Supplementary motor area; PoCG: Postcentral gyrus; SOG: Superior occipital gyrus; PCG: Posterior cingulate gyrus; CAL: Calcarine sulcus; ORBsupmed: Orbitofrontal cortex, superior medial part; IFGo: Inferior frontal gyrus, orbital part; VERM10: Vermis_10; CRBL6: Cerebellum_6.

The conventional ML analysis methods are FC-based, i.e., the FC matrices are the input of the ML methods instead of the original rs-fMRI series. This requires an additional step of data pre-processing, which needs additional time and computation, while it may also cause some loss in the functional information. Our proposed method directly uses the rs-fMRI series as the input, which successfully preserved brain functional information while reducing noise and redundancy. The input gate, forget gate, and output gate in the LSTM cell are designed to handle temporal dependencies in relatively long sequences, which could successfully extract temporal information along the time dimension in a time- and computation- efficient way.

The brain connectivity analysis of the LSTM model was based on the top ROIs with high magnitudes of the learned weights, which showed its superiority compared with the conventional whole-brain FC analysis. While the results of the LSTM ROI analysis showed similar trends as the traditional whole-brain FC analysis in terms of highlighting similar regions with increased or decreased connectivity, the very large number of connections in the whole-brain FC analysis not only hinders interpretation but also did not survive correction, thus, were not informative. Furthermore, given the relatively smaller number of subjects, the whole-brain FC analysis could result in overfitting and also be computationally expensive. The post hoc regression results of the LSTM model using the FC between the top ROIs also highlighted the connections that may be implicated in disease severity, whereas the traditional FC analysis did not show a significant relationship between disease severity and whole-brain FC across the entire group. In conclusion, the LSTM model was able to accurately classify the two early disease stages by identifying the specific brain regions and edges that contribute strongly to disease stage and motor impairment.

The current work focuses on the usage of only rs-fMRI data in early stage PD classification. Future work should introduce multimodal clinical data into the characterization, such as task-based fMRI data, cognitive and behavioral assessments, and demographic information. This will fully utilize each patient's clinical profile and more comprehensively assess factors that may predict disease stage, potentially improving stage classification and thus producing more fruitful features for characterizing differences between the two early stages.

## V. CONCLUSION

We proposed an LSTM model for early stage PD characterization using rs-fMRI data from the majority of the PPMI dataset. Under the repeated stratified cross-validation, the LSTM model significantly outperformed the other traditional ML methods in accuracy, F1 score, and recall with the highest correlation between model output scores and the MDS-UPDRS-III motor scores. The LSTM model interpretation results suggested the highly influential ROIs in early stage PD progression, and the brain connectivity analysis results identified prominent edges in brain FC changes and the disease severity. The propounded regions and edges are related to the symptoms of PD, which supports the validity of our proposed LSTM model for early stage characterization. Identification of brain regions and connectivities affected by early PD progression could potentially help unravel the mechanisms of PD and facilitate the development of new therapeutic targets.

## APPENDIX

A conventional whole-brain analysis was performed to determine FC differences between stage 1 and stage 2 on the 6,670 edges in total. Similarly to the LSTM ROI analysis, significance of the FC differences were assessed using the permutation test of the Welch's t-test, conducted on all the edges of the whole brain ROIs with 10,000 random permutations. For each edge, the significance of the Welch's t-statistic was assessed to compare whether there was a significant difference in the FC for the ROI pair between the stage 1 and stage 2 subjects. FDR correction was also applied for the large number of multiple comparisons. The significantly different edges of the whole-brain analysis (uncorrected) are visualized in Figure A1, yet none of the detected edges survived the FDR correction with the same false discovery rate of 0.2 as in the top ROI analysis. This potentially indicates that the whole-brain analysis results in overfitting and might not be reliable in revealing FC changes related to the disease stage.

An elastic net regression model was also estimated to regress the MDS-UPDRS-III motor scores onto the set of whole-brain connections with the regularization parameters set to $\alpha = 0.1744$ and l1_ratio = 1.0000. The searching range of both $\alpha$ and l1_ratio is [0, 1], the same as the regression analysis for LSTM model interpretation. The optimal parameters were selected under the same repeated cross-validation splitting strategy (3 runs, 10 folds) as the top FC analysis. Similar to the LSTM ROI analysis, significance of the regression coefficients was assessed using the permutation test. Permutation testing with 10,000 runs was conducted and the p-values were calculated as the percentage of permutation results with a coefficient magnitude greater than the magnitude of the original observation. In Figure A2, the edges with significant weights for the whole-brain regression are visualized. Note that none of the detected edges were significant after applying the same 0.2 threshold for FDR correction.

## ACKNOWLEDGMENT

Xueqi Guo was supported by the Biomedical Engineering Ph.D. fellowship from Yale University. Sule Tinaz was supported by the National Institute of Neurological Disorders and Stroke (grant number K23NS099478). We thank the Yale Center for Research Computing for use of the research computing infrastructure, specifically the Farnam cluster.

Data used in the preparation of this article were obtained from the Parkinson's Progression Markers Initiative (PPMI) database (www.ppmi-info.org/data). For up-to-date information on the study, visit www.ppmi-info.org. PPMI – a public-private partnership – is funded by the Michael J. Fox Foundation for Parkinson's Research funding partners 4D Pharma, Abbvie, Acurex Therapeutics, Allergan, Amathus Therapeutics, ASAP, Avid Radiopharmaceuticals, Bial Biotech, Biogen, BioLegend, Bristol-Myers Squibb, Calico, Celgene, Dacapo Brain Science, Denali, The Edmond J. Safra Foundaiton, GE Healthcare, Genentech, GlaxoSmithKline, Golub Capital, Handl Therapeutics, Insitro, Janssen Neuroscience, Lilly, Lundbeck, Merck,

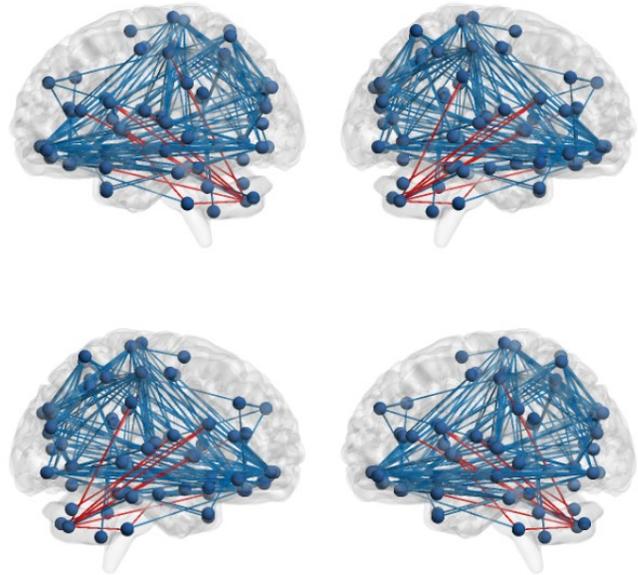

Fig. A1. The significantly different edges in stage 1 and 2 by permutation Welch's t-test for the whole brain analysis. Red: increased edges. Blue: decreased edges.

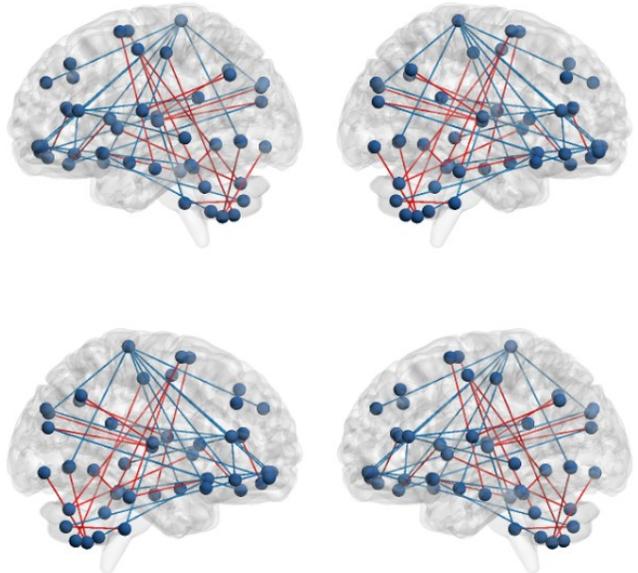

Fig. A2. The edges with significant weights by permutation t-test of elastic net regression of MDS-UPDRS-III score for the whole brain analysis. Red: increased edges. Blue: decreased edges.